\pdfoutput=1

\documentclass[11pt,twoside,a4paper,cmspaper,final,collab]{cms-tdr}
\def\svnVersion{41719M}\def\cmsCernNo{2010-053}\def\cmsCernDate{\today}\def\cmsMessage{Submitted to Physical Review Letters}

\begin{document}\cmsNoteHeader{QCD-10-019}

\hyphenation{had-ron-i-za-tion}
\hyphenation{cal-or-i-me-ter}
\hyphenation{de-vices}
\RCS$Revision: 24247 $
\RCS$HeadURL: svn+ssh://alverson@svn.cern.ch/reps/tdr2/papers/QCD-10-019/trunk/QCD-10-019.tex $
\RCS$Id: QCD-10-019.tex 24247 2010-12-01 21:25:19Z adavidzh $
\providecommand{\pthat}{\ensuremath{\hat{p}_\mathrm{T}}\xspace}
\providecommand{\ptjet}{\ensuremath{p_{\mathrm{T,jet}}}\xspace}
\providecommand{\ecal}{\textsc{ecal}\xspace}
\providecommand{\hcal}{\textsc{hcal}\xspace}
\providecommand{\dr}{\ensuremath{\Delta}R\xspace}
\providecommand{\dR}{\ensuremath{\Delta}R\xspace}
\newcommand{\etfull}{\ensuremath{E_T^\mathrm{full}}}
\newcommand{\hltlo}{{\tt HLT\_Photon15\_L1R}}
\newcommand{\hlthi}{{\tt HLT\_Photon25\_L1R}}
\newcommand{\hltlonelo}{{\tt HLT\_L1\_SingleEG5}}
\newcommand{\hltlonehi}{{\tt HLT\_L1\_SingleEG8}}
\newcommand{\lonelo}{{\tt L1\_SingleEG5}}
\newcommand{\lonehi}{{\tt L1\_SingleEG8}}
\newcommand{\eff}{\ensuremath{\epsilon_\mathrm{trig}}}

\providecommand{\HCAL} {\textsc{hcal}\xspace}

\newcommand{\ecaliso}{\ensuremath{{\rm Iso}_\mathrm{ECAL}}\xspace}
\newcommand{\hcaliso}{\ensuremath{{\rm Iso}_\mathrm{HCAL}}\xspace}
\newcommand{\trkiso}{\ensuremath{{\rm Iso}_\mathrm{TRK}}\xspace}
\newcommand{\Iso}{\ensuremath{{\rm Iso}}\xspace}
\newcommand{\see}{\ensuremath{\sigma_{\eta\eta}}\xspace}

\newcommand{\needsref}{\fixme{Needs Ref!}\xspace}
\newcommand{\pythia}{\textsc{Pythia}\xspace}
\newcommand{\jetphox}{\textsc{Jetphox}\xspace}
\newcommand{\geant}{\textsc{Geant}\xspace}
\newcommand{\minuit}{\textsc{Minuit}\xspace}
\newcommand{\etgamma}{\ensuremath{\et^\gamma}\xspace}
\newcommand{\etagamma}{\ensuremath{\eta^\gamma}\xspace}

\newcommand{\sshape}{\ensuremath{\mathcal{S}}\xspace}
\newcommand{\bshape}{\ensuremath{\mathcal{B}}\xspace}
\newcommand{\nb}{\ensuremath{N_\bshape}\xspace}
\newcommand{\ns}{\ensuremath{N_\sshape}\xspace}

\newlength{\figurewidth}
\ifthenelse{\boolean{cms@external}}{
\setlength{\figurewidth}{\columnwidth}
}{
\setlength{\figurewidth}{0.7\columnwidth}
}

\cmsNoteHeader{QCD-10-019} 
\title{Measurement of the Isolated Prompt Photon Production Cross Section in
$pp$ Collisions at $\sqrt{s}~=~7$~TeV}

\date{\today}

\abstract{
The differential cross section for the inclusive production
of isolated prompt photons has been measured as a function of the photon
transverse energy $E_\mathrm{T}^{\gamma}$ in $pp$ collisions at
$\sqrt{s}=7$~TeV using data recorded by the CMS detector at the LHC.
The data sample corresponds to an integrated luminosity of 2.9~pb$^{-1}$.
Photons are required to have a pseudorapidity $|\eta^{\gamma}|<1.45$ and
$E_\mathrm{T}^{\gamma}>21$~GeV, covering the
kinematic region $0.006 < x_\mathrm{T} < 0.086$.
The measured cross section is found to be in
agreement with next-to-leading-order perturbative QCD calculations.
}

\hypersetup{%
pdfauthor={CMS Collaboration},%
pdftitle={Measurement of the Isolated Prompt Photon Production Cross Section in
pp Collisions at 7 TeV}, pdfsubject={CMS}, pdfkeywords={CMS, physics,
isolated prompt photons}}

\maketitle 

The measurement of isolated prompt photon production in proton-proton ($pp$) collisions
provides a test of perturbative quantum chromodynamics (pQCD) and the
possibility to constrain the parton distribution functions (PDF) of the proton.
Such a measurement complements deep-inelastic scattering, Drell-Yan pair
production, and jet production measurements~\cite{ct10,nnpdf20,mstw2008}.
At the Large Hadron Collider (LHC)~\cite{LHCmachine}, a significant increase of
centre-of-mass energy with respect to previous
collider
experiments~\cite{ISRphoton,1984photonreview,SppSphoton,d0photon,cdfphoton}
allows the exploration of new kinematic regions in the hard-interaction processes in hadron-hadron collisions~\cite{DdE}.
Isolated prompt photon production also represents a background to searches
for new phenomena involving photons in the final state. 

In high-energy $pp$ collisions, single prompt photons are produced
directly in $qg$ Compton scattering and $q\bar{q}$ annihilation, and in the fragmentation of
partons with large transverse momentum.
Photons are also produced in the decay of hadrons, which can mimic prompt
production.
This background, mostly from the decays of energetic $\pi^0$ and $\eta$ mesons,
can be reduced by imposing isolation criteria on the photon candidates.

This Letter presents a measurement of the differential production cross
section of isolated prompt photons as a function of the photon transverse
energy \etgamma in $pp$ collisions at $\sqrt{s}=7~\TeV$. The analyzed data
sample corresponds to $2.9\pm0.3$~\pbinv of integrated luminosity, as recorded
by the CMS detector at the LHC~\cite{ref:EWK-10-004}.
Isolated prompt photons with a pseudorapidity $|\etagamma|<1.45$ and  $\etgamma>21~\GeV$ are studied.
Here, $\etagamma=-\ln[\tan(\theta/2)]$ and $\etgamma=E^{\gamma}\,\sin(\theta)$,
where $E^{\gamma}$ is the photon energy and $\theta$ is the polar angle of the
photon momentum measured with respect to the counterclockwise beam direction.
This measurement exploits the
difference between the electromagnetic shower profiles of prompt photons and of
photon pairs from neutral-meson decays.

Photons are detected in the
lead tungstate (PbWO$_4$) crystal electromagnetic calorimeter (ECAL), covering
$|\eta|<3.0$, comprising barrel and end cap sections.
The analysis presented in this Letter is restricted to the barrel section,
which covers $|\eta|<1.479$.
Light produced in the crystals is read out by avalanche
photodiodes (APD) in the ECAL barrel.
The ECAL barrel granularity is $\Delta\eta\times\Delta\phi=0.0174\times 0.0174$,
where $\phi$ is the azimuthal angle measured with respect to the
beam direction.
The ECAL has an ultimate energy resolution better than 0.5\% for
unconverted photons with transverse energies above $100
\GeV$~\cite{ref:ECAL-en-res}.
In 2010 collision data, for $\et>20 \GeV$, this resolution is already better
than 1\% in the barrel~\cite{ref:EGM-10-003}.
Surrounding the ECAL there is a
brass and scintillator sampling hadron calorimeter (HCAL), covering
$|\eta|<3.0$.
For $|\eta|<1.479$, the calorimeter modules are arranged in projective towers
with a segmentation of $\Delta\eta\times\Delta\phi=0.0870\times 0.0870$.
The ECAL and HCAL surround a tracking system with multiple silicon
pixel and microstrip layers, covering $|\eta|<2.5$.
Both the tracker and the calorimeters are immersed in a 3.8~T axial
magnetic field.
A detailed description of the CMS detector can be found in Ref.~\cite{cmsjinst}.

Photons are reconstructed from clusters of
energy deposited in the ECAL, using the same algorithm and granularity
at the trigger level and in the offline analysis.
Energy deposits within $|\Delta\phi|<0.304$ and
$|\Delta\eta|<0.044$ are grouped into clusters~\cite{PTDR1}.
The clustering algorithm efficiently reconstructs the energy of photons that
convert in the material in front of the ECAL.
The clustered energy is corrected taking into account
interactions in the material in front of the ECAL and electromagnetic shower
containment~\cite{ref:EGM-10-005};
the correction is parametrised as a
function of cluster size, $\eta$, $\et$, and is on average $1\%$.
The triggers used to collect the analysed data sample require
the presence of at least one reconstructed electromagnetic cluster with a minimum
transverse energy of 20 or $25~\GeV$.
The trigger is fully efficient for $\etgamma > 21\GeV$ and
$|\eta^{\gamma}|<1.45$, defining the phase space of the measurement.
Depending on the LHC instantaneous luminosity, rate-reduction factors were
applied to the triggers at 20~\GeV.
Consequently, photons with $\etgamma < 26~\GeV$ are taken from a restricted
data-set having an integrated luminosity of $2.1\pm0.2~\pbinv$.
No photon isolation criteria are applied at the trigger level.

The event selection requires at least one reconstructed primary interaction vertex
	consistent with a $pp$ collision~\cite{vertex}.
The time of the ECAL signals is required to be
compatible with that of collision products~\cite{CRAFT-ECAL-timing}.
Topological selection criteria are used to suppress direct interactions
in the ECAL APDs~\cite{ref:NOTE-2010/012}.
The residual contamination has an
effect smaller than $0.2\%$ on the measured cross section over the entire
\etgamma range considered.
Contamination from non-collision backgrounds is estimated to be
negligible~\cite{ref:EGM-10-005}.

Photon candidates are built from ECAL energy clusters fully contained in the
barrel section.
The photon candidate pseudorapidity is corrected for the position of the
primary interaction vertex.
The absolute photon energy scale is determined with electrons from
reconstructed Z-boson decays with an uncertainty estimated
to be less than $1\%$. 
Consistent results are obtained with low-energy photons from $\pi^{0}$ decays.
The linearity of the response of detector and electronics has been measured
with laser light and test beams, to a precision better than $1\%$
in the energy range probed in this Letter~\cite{ref:EGM-10-003}.
Showers initiated by charged hadrons are rejected by requiring
$E^\mathrm{HCAL}/E^{\gamma}<0.05$, where $E^\mathrm{HCAL}$ is the sum of energy
in the HCAL towers within $R<0.15$, with
$R^{2}=(\eta-\etagamma)^{2}+(\phi-\phi^{\gamma})^{2}$.
Electrons are rejected by requiring the absence of hits in the first two layers
of the pixel detector that would be consistent with an electron track matching
the observed location and energy of the photon candidate (pixel veto
requirement).

The photon candidates must satisfy three isolation requirements that reject
photons produced in hadron decays:
(1) $\trkiso<2~\GeVc$, where \trkiso is the sum of the \pt of tracks
compatible with the primary event vertex in an annulus $0.04<R<0.40$,
excluding a rectangular strip of $\Delta\eta\times\Delta\phi=0.015\times0.400$
to remove the photon's own energy if it converts into an $e^{+}e^{-}$ pair;
(2) $\ecaliso<4.2~\GeV$, where \ecaliso is the transverse energy deposited in
the ECAL in an annulus $0.06<R<0.40$, excluding a rectangular strip of
$\Delta\eta\times\Delta\phi=0.04\times0.40$; and
(3) $\hcaliso<2.2~\GeV$, where \hcaliso is the transverse energy deposited in
the HCAL in an annulus $0.15<R<0.40$.
The requirements were designed with two other objectives in mind.
First, the use of relatively loose photon identification and isolation selection
criteria reduces the dependence of the results on the details of the simulation
of the detector noise, the underlying event, and event pile-up.
Second, the shape of the isolation regions is designed to allow the
use of electrons to determine the efficiency of the isolation requirements in data.
The isolation requirements also reduce the uncertainty on the signal
due to the knowledge of the photon fragmentation functions.
In total, $4\times10^{5}$ photon candidates fulfill the selection criteria.

While the isolation requirements remove the bulk of the neutral-meson
background, a substantial contribution remains, mainly caused by fluctuations
in the fragmentation of partons, where neutral mesons carry most of the
energy and are isolated.
A modified second moment of the electromagnetic energy cluster about its mean
$\eta$ position, \see, is used to measure the isolated
prompt photon yield.
It is calculated as $$\sigma^2_{\eta\eta} =
\sum_{i=1}^{25}w_i(\eta_{i}-\bar{\eta})^2 / \sum_{i=1}^{25}w_i,$$ where
$w_i = \max(0, 4.7+\ln(E_i/E)),$
$E_{i}$ is the energy of the $i^\mathrm{th}$ crystal in a
group of $5\times5$ centred on the one with the highest energy, and
$\eta_{i}=\hat{\eta}_{i}\times\delta\eta$, where $\hat{\eta}_{i}$ is the $\eta$
index of the $i$th crystal and $\delta\eta=0.0174$; $E$ is the total
energy of the group and $\bar{\eta}$ the average $\eta$ weighted by $w_i$ in
the same group~\cite{positionLog}.
Since \see expresses the extent in $\eta$ of the cluster, it
discriminates between clusters belonging to isolated prompt photons, for which
the \see distribution is very narrow and symmetric, and
photons produced in hadron decays, for
which the distribution is dominated by a long tail towards higher values.
Given the axial configuration of
the CMS magnetic field, interactions with the material in front of the ECAL have
a small influence on the shower profile along the $\eta$ direction,
such that \see is not affected by uncertainties on the modeling of such effects.
The mean of the \see distributions is found to be independent of the number of
reconstructed interaction vertices, and therefore it does not show sensitivity
to pileup interactions.

\begin{figure}[htbp]
	\begin{center}
        \includegraphics[width=\figurewidth
        ]{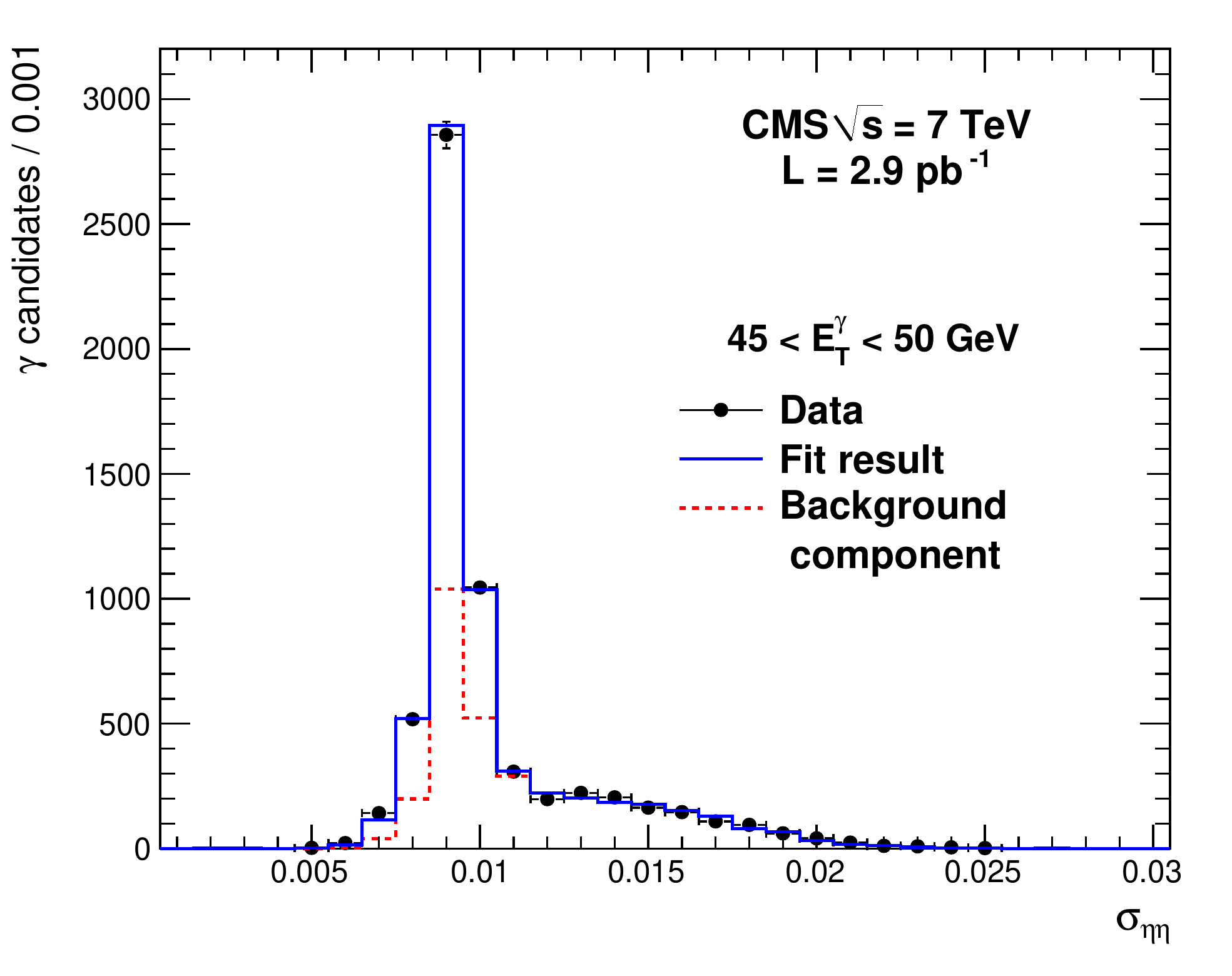}
        \caption{
          Measured \see distribution for photons with $45<\etgamma<50~\GeV$.
          The fit result (solid) and the background component
          (dashed) are also shown.
          \label{fig:fitexample}
        }
	\end{center}
\end{figure}

The isolated prompt photon yield is estimated with a binned
extended maximum likelihood fit to the \see distribution with the expected signal
and background components.
This is performed in each \etgamma bin using \minuit~\cite{minuit}.
The signal component shape is obtained from photon events
generated with \pythia~6.420~\cite{pythia64} and the
D6T parameter set~\cite{pythia-d6t}, and simulated with \geant~4~\cite{geant4}.
The \see distribution of electrons
from Z-boson decays is observed to be shifted when comparing data and simulated
events.
The shift is $+(8\pm3)\times10^{-5}$ and corresponds to $0.9\%$ of the
average of the simulated photon \see values, which are corrected for the
observed shift.
The background component shape is extracted from data by taking the \see
distribution of events in a background-enriched isolation sideband defined by
requiring $2<\trkiso<5~\GeVc$, while keeping all other selection criteria
unchanged.
This choice provides a sufficient number of events while minimising the bias to
the \see distribution due to the positive correlation between \see and
\trkiso.
Both signal and background shapes are obtained separately for each \etgamma bin.
Figure~\ref{fig:fitexample} illustrates the result of the two-component fit for
the $45<\etgamma<50~\GeV$ bin, which is representative of the fits in all
\etgamma bins.
The isolated prompt photon signal yield, $N^\gamma$, is extracted with this
fitting procedure.
For $\see<0.01$, the fraction of isolated prompt photons in data after full
selection increases from $38\%$ at $\etgamma=25~\GeV$ to $80\%$ at
$\etgamma=100~\GeV$.

The  differential cross section as a function of \etgamma is
defined as
$$d^{2}\sigma/d\etgamma d\etagamma = N^\gamma /
(L\;\mathcal{U}\;\epsilon\;\Delta\etgamma\;\Delta\etagamma),$$
where $\Delta\etgamma$ is the size of the $\etgamma$ bin,
$\Delta\etagamma=2.9$, $L$ is the integrated luminosity, and $\mathcal{U}$
denotes bin-by-bin corrections that account for \etgamma reconstruction effects
and finite detector resolution in \etagamma and isolation quantities.
The overall efficiency $\epsilon$ is the
product of the photon trigger, reconstruction, and selection efficiencies.
The trigger is fully efficient for $\etgamma > 21\GeV$ and
$|\eta^{\gamma}|<1.45$, as previously mentioned.
The efficiency of the isolation criteria is measured in data using an
electron sample from Z-boson decays and is found to be
higher than in simulation by
$\rho_\epsilon=1.035\pm0.017\,\mathrm{(stat+syst)}$.
The photon reconstruction and selection efficiencies are determined from \pythia
events with prompt photons and are scaled by $\rho_\epsilon$.
The estimated efficiency is $\epsilon=0.916\pm0.034\,\mathrm{(stat+syst)}$ and
does not change appreciably with $\etgamma$ or $\etagamma$.
Using events generated with \pythia, the values of $\mathcal{U}$ are calculated
as a function of \etgamma for prompt photons with $|\etagamma|<1.45$ and
particle-level isolation less than $5~\GeV$.
The latter is defined as the sum of the \pt of simulated
particles within $R<0.4$.
The resulting values of $\mathcal{U}$ decrease from $1.01$ to $0.97$ as \etgamma
increases and are listed in Table~\ref{tab:data_theory}.

The total systematic uncertainty on the measured cross section includes
contributions from the uncertainties in
the shapes of the \see distributions of signal and background,
the efficiency,
the photon energy scale,
the binning of the \see distributions, and
the $\mathcal{U}$ corrections.
The largest contribution is due to the limited knowledge
of the background component shape, which affects the measurement for two
reasons.
First, photon candidates selected from the isolation sideband have more
associated activity in the isolation region than the true
background.
This effect is investigated by comparing the sideband and true \see
distributions in simulated di-jet events.
Events from the sideband emphasize the tail of the background \see
distribution, such that the cross section values extracted using the true
background \see distribution are systematically lower by $15\%$ for
$\etgamma<85~\GeV$ and $7\%$ otherwise.
Second, the sideband requirements also select some prompt photons.
This effect is investigated by comparing the isolation sideband \see
distributions of simulated samples with and without prompt photons.
Samples with prompt photons enhance the peaking part of the background
distribution, such that the cross section values extracted using the
samples without prompt photons are systematically higher by $12\%$.
These two effects are checked with data by changing the isolation sideband
limits so as to accentuate each of them.
The observed variations in the extracted cross section agree with the estimated
systematic uncertainties given above.
The systematic uncertainty on the cross section due to the efficiencies is
$\pm3.8\%$, independent of \etgamma and is dominated by the
uncertainty in the efficiency of the pixel veto requirement.
The full inefficiency of the pixel veto requirement, estimated with simulated
events, is assigned to the systematic uncertainty and is mostly due to the
rejection of prompt photons that convert in, or before, the first layer of the pixel detector.
The use of simulation to estimate this inefficiency is supported by the
$10\%$ accuracy with which the material
distribution is known~\cite{ref:TRK-10-003}.
All the other sources of uncertainty have an effect
on the measured cross section smaller than $\pm3\%$.

\begin{table}[htbp]
\centering
\caption{
Isolated prompt photon cross
section for $|\etagamma |<1.45$
and in bins of \etgamma.
Uncertainties in the cross sections are statistical.
An additional $11\%$ luminosity uncertainty is not included in the systematic
uncertainty (third column).
The last column reports the corrections
for finite detector resolution.
A correction to account for extra activity ($C=0.97\pm 0.02$) is applied
to the theoretical predictions, as explained in the text.}
\label{tab:data_theory}
\begin{tabular}{rrcc}
    \hline\hline
     \etgamma (\GeV) & $d^{2}\sigma/d\etgamma d\etagamma$
     (nb/GeV) & Syst. Unc. (\%) & $\mathcal{U}$
     \\
     \hline
  21--23   &  $        2.17 \pm0.03   $ 			   			& $+13,-16$ & 1.01 \\ 
  23--26   &  $        1.39 \pm0.02   $ 						& $+13,-16$ & 1.01 \\ 
  26--30   &  $        0.774 \pm0.010 $ 						& $+13,-16$ & 1.01 \\ 
  30--35   &  $        0.402 \pm0.006 $ 						& $+13,-16$ & 1.00 \\ 
  35--40   &  $        0.209 \pm0.004 $ 						& $+13,-16$ & 1.00 \\ 
  40--45   &  $ \left( 124.4 \pm2.8   \right) \times10^{-3}$	& $+13,-16$ & 1.00 \\
  45--50   &  $ \left( 74.0 \pm2.1    \right) \times10^{-3}$	& $+13,-16$ & 1.00 \\ 
  50--60   &  $ \left( 40.3 \pm1.0    \right) \times10^{-3}$	& $+13,-16$ & 1.00 \\ 
  60--85   &  $ \left( 12.36 \pm0.35  \right) \times10^{-3}$	& $+14,-16$ & 0.99 \\ 
  85--120  &  $ \left( 2.43 \pm0.12   \right) \times10^{-3}$	& $+14,-9 $ & 0.98 \\ 
  120--300 &  $ \left( 0.188 \pm0.013 \right) \times10^{-3}$	& $+13,-9 $ & 0.97 \\ 	
 	\hline\hline
\end{tabular}
\end{table}	

\begin{figure}[htbp]
	\begin{center}
        \includegraphics[width=\figurewidth,
        angle=0]{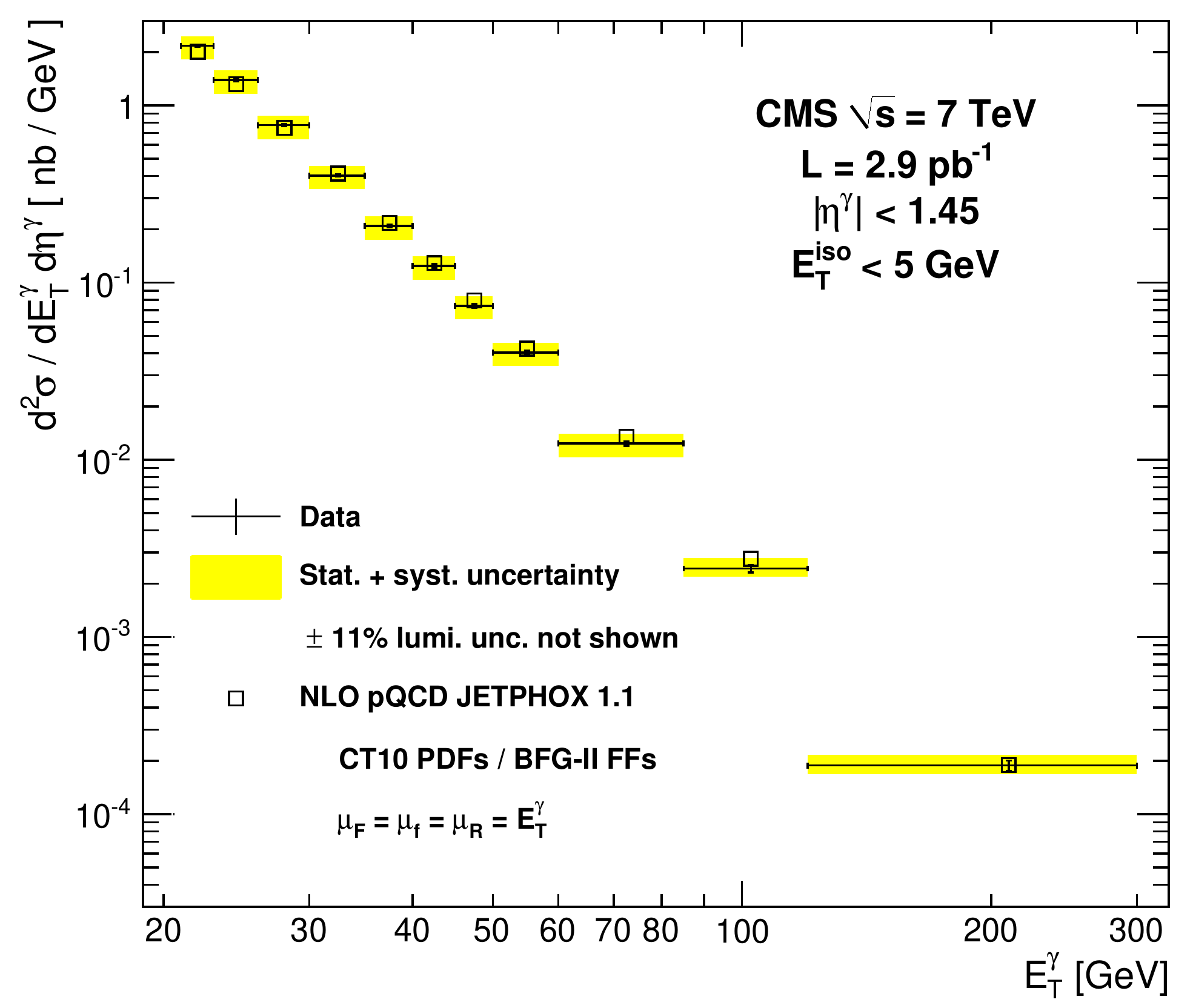}
        \caption{
        Measured isolated prompt photon differential cross section and NLO pQCD
        predictions, as a function of \etgamma.
        The vertical error bars show the statistical uncertainties, while the
        shaded areas show the statistical and systematic
        uncertainties added in quadrature.
        A correction to account for extra activity ($C=0.97\pm 0.02$) is
        applied to the theoretical predictions, as explained in the text.
		The $11\%$ luminosity uncertainty on the data is not included.
}
		\label{fig:data}
	\end{center}
\end{figure}

The measured isolated prompt photon cross section as a function of \etgamma,
including both statistical and total systematic uncertainties, is reported in
Table~\ref{tab:data_theory}.
The $11\%$ overall uncertainty on the integrated luminosity is considered
separately.
The data are shown in Fig.~\ref{fig:data}, together with next-to-leading order
(NLO) pQCD predictions from \jetphox~1.1~\cite{jetphox} using the CT10
PDFs~\cite{ct10} and the BFG set II of fragmentation functions (FF)~\cite{BFG}.
The renormalization, factorization, and fragmentation scales ($\mu_{R}$,
$\mu_{F}$, and $\mu_{f}$) are all set to \etgamma.
The hadronic energy surrounding the photon is required to be
at most $5~\GeV$ within $R<0.4$ at the parton level.
To estimate the effect of the choice of theory scales in the predictions, the
three scales are varied independently and simultaneously between $\etgamma/2$
and $2\,\etgamma$.
Retaining the largest variations the predictions change by
$(+30,-22)\%$ to $(+12,-6)\%$ with increasing \etgamma.
The uncertainty on the predictions due to the PDFs is estimated from the
envelope of predictions obtained using three global-fit parametrizations,
CT10, MSTW2008~\cite{mstw2008}, and NNPDF2.0~\cite{nnpdf20}, as recommended by
the PDF4LHC working group~\cite{pdf4lhc}.
This uncertainty is about $\pm6\%$ over the considered \etgamma range.
Predictions obtained using the CTEQ6.1M PDFs \cite{cteq6}, extensively
used in previous comparisons with data, are consistent with those obtained with
CT10 to within $3\%$.
Finally, using the BFG set I of FFs instead of the BFG set II
yields negligible differences in the predictions.
The theoretical predictions include an additional correction factor
$C(\etgamma)$ to account for the presence of contributions from the
underlying event and parton-to-hadron fragmentation, which tend to
increase the energy in the isolation cone.
Using simulated \pythia events, $C$ is determined as
the ratio between the isolated fraction of the total prompt photon cross section
at the hadron level and the same fraction obtained after turning off both
multiple-parton interactions and hadronization.
Four different sets of \pythia parameters (Z2~\cite{pythia-z2}, D6T, DWT, and
Perugia-0~\cite{pythia-p0}) are considered.
The value $C=0.97\pm 0.02$ is taken as the correction, its uncertainty covering
the results obtained with the different \pythia parameter sets.

As expected, the correction reduces the predicted cross section, since the
presence of extra activity results in some photons failing the
isolation requirements.

\begin{figure}[htbp]
	\begin{center}
        \includegraphics[width=\figurewidth,
        angle=0]{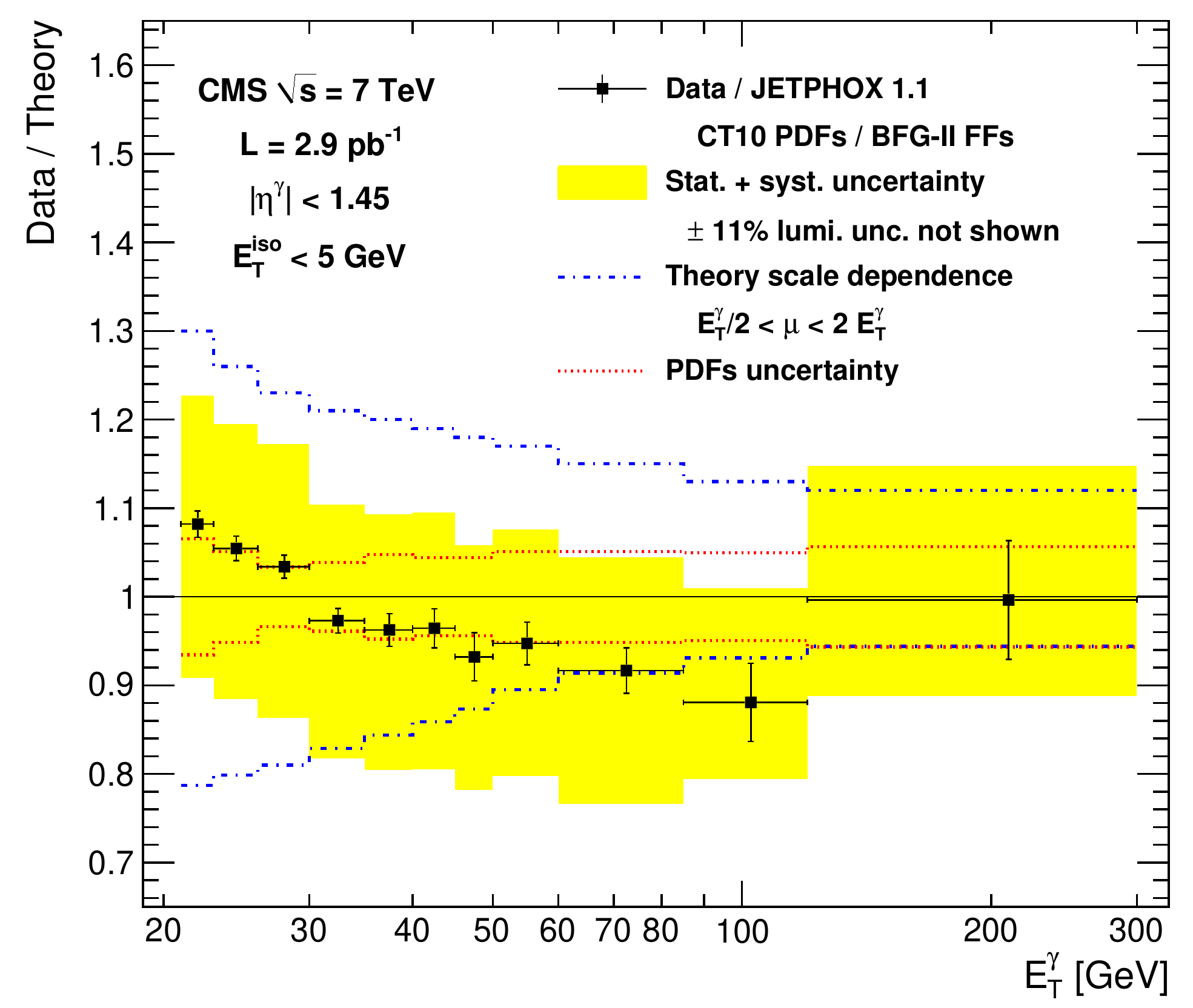} \caption{
        Ratio of the measured isolated prompt photon differential cross section
        to the NLO pQCD predictions.
        The vertical error bars show the statistical uncertainties, while the
        shaded areas show the statistical and systematic
        uncertainties added in quadrature.
        The $11\%$ luminosity uncertainty on the data is not included. The two
        sets of curves show the uncertainties on the theoretical predictions
        due to their dependency on the renormalization, factorization, and
        fragmentation scales, and on the PDFs.
        A correction to account for extra activity ($C=0.97\pm 0.02$) is
        applied to the theoretical predictions, as explained in the text. }
		\label{fig:data_theory}
	\end{center}
\end{figure}

Predictions from NLO pQCD are found to be in good agreement with
the measured cross sections, as shown in Figs.~\ref{fig:data}
and~\ref{fig:data_theory}.
The measured pattern is better described by the
theoretical predictions than in previous measurements at lower $\sqrt{s}$ and
 higher $x_\mathrm{T} = 2\ET/\sqrt{s}$
\cite{cdf1992, d01993, cdf1994, d01996, d01999, cdf2004, d0photon, cdfphoton}.

In conclusion, a measurement of the differential cross section for the production of
isolated prompt photons with $21<\etgamma<300~\GeV$ and $|\etagamma|<1.45$ in
$pp$ collisions at $\sqrt{s}=7~\TeV$ has been presented.
This measurement is performed in the kinematic regime
$0.006<x_\mathrm{T}<0.086$, probing a previously unexplored region at low
$x_\mathrm{T}$, and agrees with NLO pQCD predictions
in the whole $x_\mathrm{T}$ range.
This measurement establishes a benchmark for photon identification and
background estimation, and constrains the rate of one of the background processes
affecting searches for new physics involving photons.

We wish to congratulate our colleagues in the CERN accelerator departments for
the excellent performance of the LHC machine. We thank the technical and
administrative staff at CERN and other CMS institutes, and acknowledge support
from: FMSR (Austria); FNRS and FWO (Belgium); CNPq, CAPES, FAPERJ, and FAPESP
(Brazil); MES (Bulgaria); CERN; CAS, MoST, and NSFC (China); COLCIENCIAS
(Colombia); MSES (Croatia); RPF (Cyprus); Academy of Sciences and NICPB
(Estonia); Academy of Finland, ME, and HIP (Finland); CEA and CNRS/IN2P3
(France); BMBF, DFG, and HGF (Germany); GSRT (Greece); OTKA and NKTH (Hungary);
DAE and DST (India); IPM (Iran); SFI (Ireland); INFN (Italy); NRF and WCU
(Korea); LAS (Lithuania); CINVESTAV, CONACYT, SEP, and UASLP-FAI (Mexico); PAEC
(Pakistan); SCSR (Poland); FCT (Portugal); JINR (Armenia, Belarus, Georgia,
Ukraine, Uzbekistan); MST and MAE (Russia); MSTD (Serbia); MICINN and CPAN
(Spain); Swiss Funding Agencies (Switzerland); NSC (Taipei); TUBITAK and TAEK
(Turkey); STFC (United Kingdom); DOE and NSF (USA).

\clearpage
\bibliography{auto_generated}   

\providecommand{\href}[2]{#2}\begingroup\raggedright\begin{thebibliography}{10}%
\makeatletter
\providecommand{\hrefCMSnoop }[0]{\@secondoftwo}%
\makeatother

\bibitem{ct10}
\hrefCMSnoop {} {H.-L. Lai, M.~Guzzi, J.~Huston{ et~al.}, ``New parton
  distributions for collider physics'',} \textit{ Phys. Rev. D} \textbf{ 82}
  (Oct, 2010) 074024, \href{http://www.arXiv.org/abs/1007.2241}{\texttt{
  arXiv:1007.2241}}.
\href{http://dx.doi.org/10.1103/PhysRevD.82.074024}{\texttt{
  doi:10.1103/PhysRevD.82.074024}}.

\bibitem{nnpdf20}
\hrefCMSnoop {} {{ NNPDF} Collaboration, ``{A first unbiased global NLO
  determination of parton distributions and their uncertainties}'',} \textit{
  Nucl. Phys.} \textbf{ B838} (2010) 136,
  \href{http://www.arXiv.org/abs/1002.4407}{\texttt{ arXiv:1002.4407}}.
\href{http://dx.doi.org/10.1016/j.nuclphysb.2010.05.008}{\texttt{
  doi:10.1016/j.nuclphysb.2010.05.008}}.

\bibitem{mstw2008}
\hrefCMSnoop {} {A.~D. Martin, W.~J. Stirling, R.~S. Thorne{ et~al.}, ``{Parton
  distributions for the LHC}'',} \textit{ Eur. Phys. J.} \textbf{ C63} (2009)
  189, \href{http://www.arXiv.org/abs/0901.0002}{\texttt{ arXiv:0901.0002}}.
\href{http://dx.doi.org/10.1140/epjc/s10052-009-1072-5}{\texttt{
  doi:10.1140/epjc/s10052-009-1072-5}}.

\bibitem{LHCmachine}
\hrefCMSnoop {} {L.~Evans and P.~Bryant, ``LHC Machine'',} \textit{ JINST}
  \textbf{ 3} (2008) S08001.
\href{http://dx.doi.org/10.1088/1748-0221/3/08/S08001}{\texttt{
  doi:10.1088/1748-0221/3/08/S08001}}.

\bibitem{ISRphoton}
\hrefCMSnoop {} {M.~Diakonou {et~al.}, ``Direct production of high pT single
  photons in pp collisions at the CERN ISR'',} \textit{ Phys. Lett.} \textbf{
  B87} (1979) 292.
  \href{http://dx.doi.org/10.1016/0370-2693(79)90985-7}{\texttt{
  doi:10.1016/0370-2693(79)90985-7}}.

\bibitem{1984photonreview}
\hrefCMSnoop {} {T.~Ferbel and W.~R. Molzon, ``Direct-photon production in
  high-energy collisions'',} \textit{ Rev. Mod. Phys.} \textbf{ 56} (1984) 181.
\href{http://dx.doi.org/10.1103/RevModPhys.56.181}{\texttt{
  doi:10.1103/RevModPhys.56.181}}.

\bibitem{SppSphoton}
\hrefCMSnoop {} {{ UA2} Collaboration, ``Direct photon production at the CERN
  pbar-p collider'',} \textit{ Phys. Lett.} \textbf{ B176} (1986) 239.
\href{http://dx.doi.org/10.1016/0370-2693(86)90957-3}{\texttt{
  doi:10.1016/0370-2693(86)90957-3}}.

\bibitem{d0photon}
\hrefCMSnoop {} {{ {D0}} Collaboration, ``{Measurement of the isolated photon
  cross section in $p \bar{p}$ collisions at $\sqrt{s}$ = 1.96 TeV}'',}
  \textit{ Phys. Lett.} \textbf{ B639} (2006) 151,
  \href{http://www.arXiv.org/abs/hep-ex/0511054}{\texttt{
  arXiv:hep-ex/0511054}}.
\href{http://dx.doi.org/10.1016/j.physletb.2006.04.048}{\texttt{
  doi:10.1016/j.physletb.2006.04.048}}.

\bibitem{cdfphoton}
\hrefCMSnoop {} {{ {CDF}} Collaboration, ``{Measurement of the inclusive
  isolated prompt photon cross section in $p\bar{p}$ collisions at $\sqrt{s}$ =
  1.96 TeV using the CDF Detector}'',} \textit{ Phys. Rev.} \textbf{ D80}
  (2009) 111106, \href{http://www.arXiv.org/abs/0910.3623}{\texttt{
  arXiv:0910.3623}}.
\href{http://dx.doi.org/10.1103/PhysRevD.80.111106}{\texttt{
  doi:10.1103/PhysRevD.80.111106}}.

\bibitem{DdE}
\hrefCMSnoop {} {R.~Ichou and D.~d'Enterria, ``{Sensitivity of isolated photon
  production at TeV hadron colliders to the gluon distribution in the
  proton}'',} \textit{ Phys. Rev.} \textbf{ D82} (2010) 014015,
  \href{http://www.arXiv.org/abs/1005.4529}{\texttt{ arXiv:1005.4529}}.
\href{http://dx.doi.org/10.1103/PhysRevD.82.014015}{\texttt{
  doi:10.1103/PhysRevD.82.014015}}.

\bibitem{ref:EWK-10-004}
\href {http://cdsweb.cern.ch/record/1279145} {{ CMS} Collaboration,
  ``Measurement of {CMS} Luminosity'',} \textit{ CMS Physics Analysis Summary}
  \textbf{ \href{http://cdsweb.cern.ch/record/1279145}{CMS-PAS-EWK-10-004}}
  (2010).

\bibitem{ref:ECAL-en-res}
\hrefCMSnoop {} {P.~Adzic {et~al.}, ``{Energy resolution of the barrel of the
  CMS electromagnetic calorimeter}'',} \textit{ JINST} \textbf{ 2} (2007)
  P04004.
\href{http://dx.doi.org/10.1088/1748-0221/2/04/P04004}{\texttt{
  doi:10.1088/1748-0221/2/04/P04004}}.

\bibitem{ref:EGM-10-003}
\href {http://cdsweb.cern.ch/record/1279350} {{ CMS} Collaboration,
  ``Electromagnetic calorimeter calibration with 7 TeV data'',} \textit{ CMS
  Physics Analysis Summary} \textbf{
  \href{http://cdsweb.cern.ch/record/1279350}{CMS-PAS-EGM-10-003}} (2010).

\bibitem{cmsjinst}
\hrefCMSnoop {} {{ CMS} Collaboration, ``{The CMS experiment at the CERN
  LHC}'',} \textit{ JINST} \textbf{ 3} (2008) S08004.
\href{http://dx.doi.org/10.1088/1748-0221/3/08/S08004}{\texttt{
  doi:10.1088/1748-0221/3/08/S08004}}.

\bibitem{PTDR1}
\href {http://cdsweb.cern.ch/record/922757} {{ CMS} Collaboration, ``CMS
  Physics Technical Design Report Volume I : Detector Performance and
  Software'',} \textit{ CMS Technical Design Report} \textbf{
  \href{http://cdsweb.cern.ch/record/922757}{CMS-TDR-008-1}} (2006).

\bibitem{ref:EGM-10-005}
\href {http://cdsweb.cern.ch/record/1279143} {{ CMS} Collaboration, ``Photon
  reconstruction and identification at $\sqrt{s}$ = 7 {TeV}'',} \textit{ CMS
  Physics Analysis Summary} \textbf{
  \href{http://cdsweb.cern.ch/record/1279143}{CMS-PAS-EGM-10-005}} (2010).

\bibitem{vertex}
\hrefCMSnoop {} {{ CMS} Collaboration, ``{CMS tracking performance results from
  early LHC Operation}'',} \href{http://www.arXiv.org/abs/1007.1988}{\texttt{
  arXiv:1007.1988}}. Published in the European Physical Journal C.
\href{http://dx.doi.org/10.1140/epjc/s10052-010-1491-3}{\texttt{
  doi:10.1140/epjc/s10052-010-1491-3}}.

\bibitem{CRAFT-ECAL-timing}
\hrefCMSnoop {} {{ CMS} Collaboration, ``{Time Reconstruction and Performance
  of the CMS Electromagnetic Calorimeter}'',} \textit{ JINST} \textbf{ 5}
  (2010) T03011, \href{http://www.arXiv.org/abs/0911.4044}{\texttt{
  arXiv:0911.4044}}.
\href{http://dx.doi.org/10.1088/1748-0221/5/03/T03011}{\texttt{
  doi:10.1088/1748-0221/5/03/T03011}}.

\bibitem{ref:NOTE-2010/012}
\href {http://cdsweb.cern.ch/record/1278160} {{ CMS} Collaboration,
  ``Electromagnetic calorimeter commissioning and first results with 7 TeV
  data'',} \textit{ CMS Note} \textbf{
  \href{http://cdsweb.cern.ch/record/1278160}{CMS-NOTE-2010-012}} (2010).

\bibitem{positionLog}
\hrefCMSnoop {} {T.~C. Awes {et~al.}, ``A simple method of shower localization
  and identification in laterally segmented calorimeters'',} \textit{ Nucl.
  Instrum. Meth.} \textbf{ A311} (1992) 130.
\href{http://dx.doi.org/10.1016/0168-9002(92)90858-2}{\texttt{
  doi:10.1016/0168-9002(92)90858-2}}.

\bibitem{minuit}
\hrefCMSnoop {} {F.~James and M.~Roos, ``Minuit: a system for function
  minimization and analysis of the parameter errors and correlations'',}
  \textit{ Comput. Phys. Commun.} \textbf{ 10} (1975) 343.
\href{http://dx.doi.org/10.1016/0010-4655(75)90039-9}{\texttt{
  doi:10.1016/0010-4655(75)90039-9}}.

\bibitem{pythia64}
\hrefCMSnoop {} {T.~Sj\"{o}strand, S.~Mrenna, and P.~Z. Skands, ``{PYTHIA} 6.4
  {P}hysics and {M}anual'',} \textit{ JHEP} \textbf{ 05} (2006) 026,
  \href{http://www.arXiv.org/abs/hep-ph/0603175}{\texttt{
  arXiv:hep-ph/0603175}}.
\href{http://dx.doi.org/10.1088/1126-6708/2006/05/026}{\texttt{
  doi:10.1088/1126-6708/2006/05/026}}.

\bibitem{pythia-d6t}
{ TeV4LHC QCD Working Group} Collaboration, \hrefCMSnoop {} {M.~G. Albrow
  {et~al.}, ``{Tevatron-for-LHC Report of the QCD Working Group}''.} 2006.
\href{http://www.arXiv.org/abs/hep-ph/0610012}{\texttt{ arXiv:hep-ph/0610012}}.

\bibitem{geant4}
\hrefCMSnoop {} {{ {GEANT 4}} Collaboration, ``{GEANT} 4---a simulation
  toolkit'',} \textit{ Nucl. Instr. Meth.} \textbf{ A506} (2003) 250.
  \href{http://dx.doi.org/10.1016/S0168-9002(03)01368-8}{\texttt{
  doi:10.1016/S0168-9002(03)01368-8}}.

\bibitem{ref:TRK-10-003}
\href {http://cdsweb.cern.ch/record/1279138} {{ CMS} Collaboration, ``Studies
  of Tracker Material in the {CMS} Detector'',} \textit{ CMS Physics Analysis
  Summary} \textbf{
  \href{http://cdsweb.cern.ch/record/1279138}{CMS-PAS-TRK-10-003}} (2010).

\bibitem{jetphox}
\hrefCMSnoop {} {S.~Catani, M.~Fontannaz, J.~P. Guillet{ et~al.}, ``Cross
  section of isolated prompt photons in hadron-hadron collisions'',} \textit{
  JHEP} \textbf{ 05} (2002) 028,
  \href{http://www.arXiv.org/abs/hep-ph/0204023}{\texttt{
  arXiv:hep-ph/0204023}}.
\href{http://dx.doi.org/10.1088/1126-6708/2002/05/028}{\texttt{
  doi:10.1088/1126-6708/2002/05/028}}.

\bibitem{BFG}
\hrefCMSnoop {} {L.~Bourhis, M.~Fontannaz, and J.~P. Guillet, ``{Q}uark and
  gluon fragmentation functions into photons'',} \textit{ Eur. Phys. J.}
  \textbf{ C2} (1998) 529,
  \href{http://www.arXiv.org/abs/hep-ph/9704447}{\texttt{
  arXiv:hep-ph/9704447}}.
\href{http://dx.doi.org/10.1007/s100520050158}{\texttt{
  doi:10.1007/s100520050158}}.

\bibitem{pdf4lhc}
\hrefCMSnoop {} {M.~Botje {et~al.}, ``{The PDF4LHC Working Group Interim
  Recommendations}''.} 2011.
\href{http://www.arXiv.org/abs/1101.0538}{\texttt{ arXiv:1101.0538}}.

\bibitem{cteq6}
\hrefCMSnoop {} {J.~Pumplin, D.~R. Stump, J.~Huston{ et~al.}, ``{N}ew
  {G}eneration of {P}arton {D}istributions with {U}ncertainties from {G}lobal
  {QCD} analysis'',} \textit{ JHEP} \textbf{ 07} (2002) 012,
  \href{http://www.arXiv.org/abs/hep-ph/0201195}{\texttt{
  arXiv:hep-ph/0201195}}.
\href{http://dx.doi.org/10.1088/1126-6708/2002/07/012}{\texttt{
  doi:10.1088/1126-6708/2002/07/012}}.

\bibitem{pythia-z2}
R.~Field, ``{Early LHC Underlying Event Data - Findings and Surprises}'', in
  \textit{ Proceeedings of the Hadron Collider Physics Symposium 2010}.
\newblock 2010.
\newblock
\href{http://www.arXiv.org/abs/1010.3558}{\texttt{ arXiv:1010.3558}}.
\newblock

\bibitem{pythia-p0}
P.~Z. Skands, ``{The Perugia Tunes}'', in \textit{ Proceedings of the 1st
  International Workshop on Multiple Partonic Interactions at the {LHC}},
  p.~284.
\newblock 2009.
\newblock
\href{http://www.arXiv.org/abs/1003.4220v1}{\texttt{ arXiv:1003.4220v1}}.
\newblock

\bibitem{cdf1992}
\hrefCMSnoop {} {{ {CDF}} Collaboration, ``{M}easurement of the isolated prompt
  photon cross-sections in $\bar{p}p$ collisions at $\sqrt{s} = 1.8$ {TeV}'',}
  \textit{ Phys. Rev. Lett.} \textbf{ 68} (1992) 2734.
\href{http://dx.doi.org/10.1103/PhysRevLett.68.2734}{\texttt{
  doi:10.1103/PhysRevLett.68.2734}}.

\bibitem{d01993}
\hrefCMSnoop {} {{ {CDF}} Collaboration, ``{P}rompt photon cross-section
  measurement in $\bar{p}p$ collisions at $\sqrt{s} = 1.8$ {TeV}'',} \textit{
  Phys. Rev.} \textbf{ D48} (1993) 2998.
\href{http://dx.doi.org/10.1103/PhysRevD.48.2998}{\texttt{
  doi:10.1103/PhysRevD.48.2998}}.

\bibitem{cdf1994}
\hrefCMSnoop {} {{ {CDF}} Collaboration, ``{P}recision {M}easurement of the
  {P}rompt {P}hoton {C}ross {S}ection in $p\bar{p}$ {C}ollisions at $\sqrt{s} =
  1.8$ {TeV}'',} \textit{ Phys. Rev. Lett.} \textbf{ 73} (1994) 2662.
\href{http://dx.doi.org/10.1103/PhysRevLett.73.2662}{\texttt{
  doi:10.1103/PhysRevLett.73.2662}}.

\bibitem{d01996}
\hrefCMSnoop {} {{ {D0}} Collaboration, ``{I}solated {P}hoton {C}ross {S}ection
  in the {C}entral and {F}orward {R}apidity {R}egions in $p\bar{p}$
  {C}ollisions at $\sqrt{s} = 1.8$ {TeV}'',} \textit{ Phys. Rev. Lett.}
  \textbf{ 77} (1996) 5011,
  \href{http://www.arXiv.org/abs/hep-ex/9603006}{\texttt{
  arXiv:hep-ex/9603006}}.
\href{http://dx.doi.org/10.1103/PhysRevLett.77.5011}{\texttt{
  doi:10.1103/PhysRevLett.77.5011}}.

\bibitem{d01999}
\hrefCMSnoop {} {{ {D0}} Collaboration, ``{I}solated {P}hoton {C}ross {S}ection
  in $p\bar{p}$ {C}ollisions at $\sqrt{s} = 1.8$ {TeV}'',} \textit{ Phys. Rev.
  Lett.} \textbf{ 84} (2000) 2786,
  \href{http://www.arXiv.org/abs/hep-ex/9912017}{\texttt{
  arXiv:hep-ex/9912017}}.
\href{http://dx.doi.org/10.1103/PhysRevLett.84.2786}{\texttt{
  doi:10.1103/PhysRevLett.84.2786}}.

\bibitem{cdf2004}
\hrefCMSnoop {} {{ {CDF}} Collaboration, ``{D}irect photon cross section with
  conversions at {CDF}'',} \textit{ Phys. Rev.} \textbf{ D70} (2004) 074008,
  \href{http://www.arXiv.org/abs/hep-ex/0404022}{\texttt{
  arXiv:hep-ex/0404022}}.
\href{http://dx.doi.org/10.1103/PhysRevD.70.074008}{\texttt{
  doi:10.1103/PhysRevD.70.074008}}.

\end{thebibliography}\endgroup

\cleardoublepage\appendix\section{The CMS Collaboration \label{app:collab}}\begin{sloppypar}\hyphenpenalty=5000\widowpenalty=500\clubpenalty=5000\textbf{Yerevan Physics Institute,  Yerevan,  Armenia}\\*[0pt]
V.~Khachatryan, A.M.~Sirunyan, A.~Tumasyan
\vskip\cmsinstskip
\textbf{Institut f\"{u}r Hochenergiephysik der OeAW,  Wien,  Austria}\\*[0pt]
W.~Adam, T.~Bergauer, M.~Dragicevic, J.~Er\"{o}, C.~Fabjan, M.~Friedl, R.~Fr\"{u}hwirth, V.M.~Ghete, J.~Hammer\cmsAuthorMark{1}, S.~H\"{a}nsel, C.~Hartl, M.~Hoch, N.~H\"{o}rmann, J.~Hrubec, M.~Jeitler, G.~Kasieczka, W.~Kiesenhofer, M.~Krammer, D.~Liko, I.~Mikulec, M.~Pernicka, H.~Rohringer, R.~Sch\"{o}fbeck, J.~Strauss, A.~Taurok, F.~Teischinger, W.~Waltenberger, G.~Walzel, E.~Widl, C.-E.~Wulz
% ...
\vskip\cmsinstskip
% ...
1:~~Also at CERN, European Organization for Nuclear Research, Geneva, Switzerland\\
% ...

\end{sloppypar}
\end{document}